 	\documentclass[useAMS,usenatbib]{mn2e}

	\usepackage[english]{babel}
	\usepackage{graphicx}
	
	\title[The WHIM and SZ effect]{The contribution of the Warm-Hot 
	Intergalactic Medium to the CMB anisotropies via the Sunyaev-Zeldovich effect}

	\author[I. F. Suarez-Vel{\'a}squez, J. P. M\"{u}cket, and F. Atrio-Barandela]{
	I. F. Suarez-Vel\'{a}squez$^{1}$\thanks{E-mail:isuarez@aip.de; 
	jpmuecket@aip.de; atrio@usal.es}, 
	J. P. M\"{u}cket$^{1}$ and F. Atrio-Barandela$^{2}$\\
	$^{1}$Leibniz-Institut f\"{u}r Astrophysik Potsdam, An der Sternwarte 16, 
	14482, Potsdam, Germany\\
	$^{2}$Fisica Teorica, Universidad de Salamanca, Plaza de la Merced s/n, 
	37008 Salamanca, Spain}
	
\begin{document}
	
	\date{Accepted . Received  ; in original form 	2012  }
	
	\pagerange{\pageref{firstpage}--\pageref{lastpage}} \pubyear{2012}
	
	\maketitle
	
	\label{firstpage}
	

\begin{abstract}
Cosmological hydrodynamical simulations
predict that a large fraction of all baryons reside within mildly 
non-linear structures with temperatures in the range $10^{5}-10^{7}$K. 
As the gas is highly ionized, it could be detected by the temperature 
anisotropies generated on the Cosmic Microwave Background radiation.
We refine our previous estimates of the thermal 
Sunyaev-Zeldovich effect by introducing a non-polytropic equation of state 
to model the temperature distribution of the shock heated gas
derived from temperature-density phase diagrams of 
different hydrodynamical simulations. Depending on the specific model,
the Comptonization parameter varies in the range $10^{-7}\le y_c
\le 2\times 10^{-6}$, compatible with the FIRAS upper limit. 
This amplitude is in agreement with a simple toy model constructed 
to estimate the average effect induced by filaments of ionized gas.
Using the log-normal probability density function we calculate the 
correlation function and the power spectrum of the temperature 
anisotropies generated by the WHIM filaments. 
For a wide range of the parameter space, the maximum amplitude
of the radiation power spectrum is $(\ell+1)\ell C_{\ell}/2\pi
=0.7-70(\mu K)^2$ at $\ell \approx 200-500$.
This amplitude scales with baryon density, Hubble constant and 
the amplitude of the matter power spectrum $\sigma_8$
as $[(\ell+1)\ell C_\ell]_{\mathrm{max}}/2\pi \propto \sigma_8^{2.6}(\Omega_b h)^2$.
Since the thermal Sunyaev-Zeldovich effect has a specific frequency
dependence, we analyze the possibility of detecting this component 
with the forthcoming Planck data. 
\end{abstract}	
	
	
	\section{Introduction}
The  distribution of baryons in the local Universe is one of the main problems 
of modern Cosmology. The highly ionized intergalactic gas, that evolved from the 
initial density perturbations into a complex network of mildly non-linear structures 
in the redshift interval $2 < z < 6$ \citep{rauch1998,stocke.shull.penton2004}, 
could contain most of the baryons in the universe 
\citep{rauch.et.al1997,schaye2001,richter.et.al2006}.  
At redshifts $z>2$ most baryons are found in the Ly$\alpha$ systems
detected through absorption lines in the spectra of 
distant quasars. With cosmic evolution, the baryon fraction in 
these structures decreases as more matter is concentrated within compact 
virialized objects. At low redshifts, the Ly$\alpha$ systems are
filaments with low HI column densities containing $\sim 30$\% of all 
baryons \citep{stocke.shull.penton2004}, while the material in stars and galaxies contains 
about $10\%$ of all baryons. An extra 5\% could be in the form of Circumgalactic
Medium around galaxies \citep{gupta2012}. More importantly,
at $z\sim 0$ about 50\% of all cosmic baryons 
have not yet been identified \citep{fukugita.peebles.2004,shull.smith.danforth2011}. 
Hydrodynamical simulations predicted that baryons could be in the
form of shock-heated intergalactic gas in mildly-nonlinear structures
with temperatures $10^5-10^7$K, called  Warm-Hot Intergalactic
Medium (WHIM). The baryon fraction in this medium could be as large as
40\%-50\% in the local Universe \citep{cen.ostriker1999,dave.hernquist.katz.weinberg1999,dave.cen.ostriker.et.al2001,danfforth.shull2008,Prochaska.Tumlinson,smith.et.al2011}
containing the bulk, if not all, the unidentified baryons.

The observational effort has concentrated in searching for the WHIM X-ray
signature both in emission and in absorption. \cite{soltan2006} looked for 
the extended soft X-ray emission around field galaxies but his task was 
complicated by the need to subtract all systematic effects that could mimic the
diffuse signal. More recently, the effort has been concentrated in searching 
for absorption lines due to highly ionized heavy elements from the far-ultraviolet 
to the soft X-ray (see \citeauthor{shull.smith.danforth2011} 
\citeyear{shull.smith.danforth2011}, for a review)
with partial success. For example, \cite{Dietrich} have recently reported 
the detection of a large-scale filament between two clusters of galaxies.

While absorption lines will identify individual systems, complementary
techniques are needed to study the overall properties of the WHIM.
To this purpose, \cite{atrio.muecket2006} and \cite{atrio.muecket.santos2008}
suggested that the WHIM distribution would generate temperature anisotropies 
on the Cosmic Microwave Background (CMB)
due to the thermal (TSZ) and kinematic (KSZ) Sunyaev-Zeldovich effect
\citep{sunyaev.zeldovich1972,sunyaev.zeldovich1980}, opening a new
observational window to study the gas distribution at low redshifts
\citep{genova.atrio.muecket2009}.
\cite{hallman.et.el2007} found that about a third of the SZ flux would be 
generated by unbound gas, but their estimated amplitude was lower than
the prediction of our model. Our computation assumed that the weakly 
non-linear matter distribution was described by a log-normal probability
distribution function PDF \citep{coles.jones1991,choudhury2011} and 
the gas followed a polytropic
equation of state. While the IGM at redshifts $z>2$ behaves like a polytrope, 
at low redshifts the polytropic equation of state is not longer valid since
the gas is heated mostly by shocks to  
temperatures about $10^5 K < T < 10^7 K$ at density contrasts  $\delta < 100$.
In the regions where the gas is shock-heated, small-scale density perturbations 
can be neglected due to the high pressure gradients \citep{klar.muecket2010}, 
simplifying the treatment. In this paper we shall improve our previous
calculations by using equations of state derived from hydrodynamical simulations
to provide a more physical description of the WHIM. Briefly,
in section \ref{estimate} we estimate 
the order of magnitude contribution of a population of WHIM filaments to 
the Comptonization parameter based on a crude 
geometrical model; in section \ref{logmodel} we describe our log-normal
model and we detail the differences with our previous treatments;
in section \ref{compton} we compute the CMB Comptonization parameter
due to the WHIM and in section \ref{spower} the correlation function
and angular power spectrum; in section \ref{results} we present our  
results and, finally, in section \ref{conclusion} we summarize our conclusions.

\section{Estimate of the mean Comptonization parameter due to the WHIM}\label{estimate}

The temperature anisotropies due to the inverse Compton scattering of CMB photons
by the free electrons have only been measured for clusters of galaxies. 
There are two contributions: the thermal component (TSZ) due to the motion of the
electrons in the potential wells and the kinematic component (KSZ) 
due to the motion of the cluster as a whole. The induced temperature anisotropies are
\begin{equation}\label{szeq}
\left(\frac{\delta T}{T_0}\right)_{\mathrm{TSZ}}=G(\nu)y,\quad 
\left(\frac{\delta T}{T_0}\right)_{\mathrm{KSZ}}=-\tau\frac{v_{cl}}{c},
\end{equation}
where $G(x)=x\coth(x/2)-4$, $x=h\nu/k_B T$ is the CMB frequency in dimensionless units,
$v_{cl}$ is the velocity of the cluster projected along the line of sight ({\it los})
and $c$ the speed of light.
The Comptonization parameter and the optical depth $\tau$ are defined as
\begin{equation}
y=y_0\int T_e\,n_e\,dl\,,\quad
\tau=\sigma_T\int n_e\,dl\,,
\end{equation}
where $T_e$ and $n_e$ are the electron temperature and electron number density,
$dl$ is the proper distance along the {\it los}
and $y_0=\sigma_T k_B/m_e c^2$, with $\sigma_T$, $k_B$, $m_e$ 
Thomson cross section, Boltzmann constant and electron mass, 
respectively. While the KSZ effect has the same frequency dependence
as the intrinsic CMB anisotropies, the change with frequency of TSZ effect, 
$G(x)$, is different from that of any other known foreground. It is negative
in the Rayleigh-Jeans and positive in the Wien regions, being null close to
$\nu = 217$GHz and can be more easily separated from other contributions than 
KSZ anisotropies. Since they are more easily detectable, here
we shall discuss only the TSZ contribution of the WHIM. 

	\begin{figure}
	\includegraphics[width=80mm]{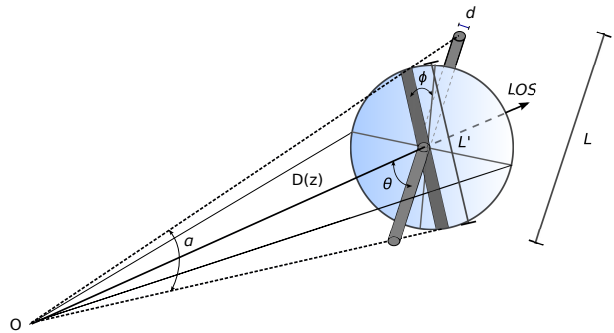}
	\caption{Geometrical model of a simple filament with constant
	density and temperature. The filament has a size $L$,
	width $d$, is located at redshift $z$ at a comoving distance $D(z)$, 
	subtends an angle $\alpha$ from the observer and is randomly oriented
	in space, forming an angle $\theta$ with respect to the line of sight.}
	\label{configuracion}
	\end{figure}	

An order of magnitude estimate of the WHIM TSZ effect can 
be obtained by considering the effect on the CMB radiation
of a single filament with constant electron density 
$n_e$ and constant temperature $T_e$. The geometrical configuration
of a filament of size $L$ and thickness $d$ is shown in 
Figure~\ref{configuracion}. For a filament randomly oriented $\omega\equiv(\theta,\phi)$ 
in space, let $\theta$ be the angle between the filament
and the {\it los}. The Comptonization parameter induced in the radiation
crossing this filament is
\begin{equation}\label{1}
y=y_0 \frac{n_e \,T_e\, d}{\cos\theta}\,,
\end{equation}
where $d/\cos\theta$ is the projected thickness of the filament.
The Comptonization parameter averaged over all possible azimuths $\phi$ is
\begin{equation}\label{2}
\langle y\rangle_{\phi} =y_0\frac{n_e \,T_e d}{\cos\theta}\frac{d}{2\pi L^{\prime}}\,.
\end{equation}	
If the size of filaments at redshift $z$ is small compared to its angular diameter distance,
$L\ll D_A(z)$, then the projected comoving length $L'$ can be written as
$L^{\prime}=L \cos\theta\sim \alpha D(z)$ where $D(z)$ is the co-moving distance 
and $\alpha$ is the angle subtended by the filament on the sky. Then,
\begin{equation}\label{6}
\langle y\rangle_{\phi} =y_0\frac{n_e \,T_e\, L d^2}{2\pi \alpha^2 D^2(z)}\,.
\end{equation}
The probability that a given $los$ crosses the filament is $L/(\pi D(z))$; 
if $d\ll L$, then the  average over all possible values of $\alpha$ is
\begin{equation}\label{7}
\langle y\rangle=y_0\frac{ n_e \,T_e}{2{\pi}^2}\frac{L}{D(z)}d\,.
\end{equation}
In their analysis of the properties of filaments,
\cite{klar.muecket2012} obtained $d=\epsilon L$, with $\epsilon \in [0.01-0.1]$.

The mean distortion along the $los$ can be obtained by adding 
the effect of all filaments of different masses at different redshifts:
\begin{equation}
\langle y\rangle=\frac{1}{2\pi}\int_z \int_{M}\frac{dn_f}{dM}\,dM\,
y_0\,n_e\,T_e\,\frac{\epsilon\,L^{2}(M)}{D(z)}\frac{d V(z)}{dz}\,dz\,,
\end{equation}
where $n_f$ is the number of filaments per unit of volume. To proceed further 
we assume that filaments connect clusters and groups of galaxies. Then, their 
length is approximately the distance between clusters. If 
$n_{cl}$ is the number of clusters of a given mass per unit of volume, then the
mean separation of the filaments within a volume $V$
is $l\sim (V/n_{cl})^{1/3}$ and its mean length $L\sim n_{cl}^{-1/3}$.
Thus, the mean Comptonization parameter along the {\it los} is
\begin{equation}
\langle y\rangle=\frac{2}{\pi}\int_z \int_{M}\frac{dn_{f}}{dM}\,dM\,
y_0 n_e T_e\epsilon \frac{D(z)}{n_{cl}(M)^{- 2/3}}\,\frac{dD(z)}{dz}\,dz\,,
\label{mean_comp}
\end{equation}
with 
\begin{equation}
dD(z)=cH_0^{-1}[\Omega_{\Lambda}+\Omega_m(1+z)^3+\Omega_k(1+z)^2]^{1/2}dz\,,
\label{dldz}
\end{equation}
and $\Omega_k=(1-\Omega_m-\Omega_{\Lambda})$.
Equation~(\ref{mean_comp}) can be used to obtain an order of magnitude of 
the distortion induced by the electrons on the WHIM  filaments. 
If we take $T_e\approx 10^7$K, $\epsilon=0.05$ and compute the number of filaments 
according to Sheth \& Tormen (2002) then $y_c=10^{-6}$, compatible with the FIRAS 
upper limit of $\bar{y}_c=1.5\times 10^{-5}$. The CMB distortion due to a 
network of filaments of ionized gas is compatible with observations.
The integration is performed from $M_i=10^{13}M_\odot$ to $M_f=10^{15}M_\odot$. 
Our results are insensitive to the exact value of $M_f$ whose number density
is exponentially suppressed but depend on $M_i$ which we took to be the mass 
of rich groups of galaxies, limit where the number of objects is not longer well
described by Sheth \& Tormen (2002). Using that formalism \citet{shimon.et.al2012} have calculated the radiation power spectrum. 
In the next section we will give a more refined prediction 
based on a log-normal model of the baryon distribution.

\section{The log-normal baryon density distribution model}
\label{logmodel}

In perturbation theory, the density contrast is defined as
\begin{equation}
\delta(\mathbf{x})\equiv 
\frac{\rho(\mathbf{x})-\langle\, \rho \,\rangle}{\langle\, \rho \,\rangle}\,,
\end{equation}
where $\rho(\mathbf{x})$ is the density at any given point and $\langle\rho\rangle$
the mean density at any given redshift. 
In Fourier space and in the linear regime, all modes evolve at the same
rate and 
\begin{equation}
\delta(\mathbf{k},z)=D_+(z)\delta(\mathbf{k},0)\,,
\end{equation}
where $D_+(z) = D_+(z,\Omega_{\Lambda},\Omega_m)$ is the linear growth factor normalized 
to $D_+(0)=1$. If the initial density field is a Gaussian random field, then the linearly 
extrapolated power spectrum $P(k)$ is defined by
\begin{equation}\label{delta-ps}
\langle \delta(\mathbf{k},0)\delta(\mathbf{k'},0)\rangle
=(2\pi)^{3}P(\mathbf{k})\delta(\mathbf{k}-\mathbf{k'})\,.
\end{equation}
In eq. \ref{delta-ps} the power spectrum $P(k)$ is that of the background
$\Lambda$CDM model, computed using linear theory. We took the
spectral index at large scales to be $n_S=1$. We did not consider corrections 
due to the non-linear evolution of the matter density field on scales 
about and below $8h^{-1}$Mpc.

In the linear regime, baryons follow the Dark Matter (DM) distribution,
but this is not longer true once structures become non-linear. In regions where 
large-scale perturbations develop shocks, \cite{klar.muecket2010} showed that 
the gas perturbations on small scales were suppressed by the
enhanced pressure and would be nearly erased. Then, the distribution of 
the baryon WHIM filaments that are forming and evolving within the large-scale 
density perturbations would be different from that of the DM. 
The linear density contrast of baryons in the IGM can be obtained from 
that of the DM by smoothing over scales below $L_0$, the largest
scale erased by shock-heating. According to \cite{fang1993}
\begin{equation}
\delta_B(\mathbf{k},z)=\frac{\delta_{\mathrm{DM}}(\mathbf{k},z)}{1+L_0^2 {k}^2}\,.
\label{power_B}
\end{equation}
The comoving length scale $L_0$ can be determined by imposing that it is the smallest 
possible length scale at which the linear peculiar velocity $v_p$ equals 
the sound speed $v_s(z)$ of the baryon fluid. The sound speed is determined 
by the mean temperature $T_{\mathrm{IGM}}(z)$ of the IGM at any given $z$:
$v_s=\sqrt{2k_B T_{\mathrm{IGM}}(z)/m_p}$. The condition $v_p \geq v_s$ gives
\begin{equation}\label{L0}
L_0(z) =\frac{2\pi (1 + z)^2\, v_s H_0^{-1}}{(\Omega_{\Lambda} + \Omega_m(1+z)^3)^{1/2}
D_+(z) \delta_0}\,.
\end{equation}
In eq.~(\ref{L0}), the sound speed is determined by the
IGM temperature $T_{\mathrm{IGM}}$ at mean density. $T_{\mathrm{IGM}} \approx 10^4 K$ and it is mainly determined by the evolution of the UV background. For that temperature, at redshift $z=0$ we have
$L_0 \approx 1.7 h^{-1}$ Mpc. This value of cut-off scale agrees with the results 
of \cite{klar.muecket2012} on the formation and evolution of WHIM filaments.
\cite{tittley2007} have studied different ionization models and discussed the possible range of variation of the mean IGM temperature.  In our calculations, we
assume that $T_{\mathrm{IGM}}$ varies as given in \cite{theuns2002}. Below we shall see that at redshifts $z > 3$ the WHIM does not
generate significant anisotropies.  At redshifts $z\le 3$, the mean IGM  
temperature variation with redshift is small and can be roughly approximated as
$\log_{10}(T_{\mathrm{IGM}}/10^3{\rm K})\approx  (A +0.1(1+z))$, where 
$A$ varies in the range $A=0.5-0.9$, corresponding to $T_{\mathrm{IGM}}=10^{3.6}-10^4$K.

The Fourier transform of eq.~(\ref{power_B}) gives the density contrast in real 
co-moving space, $\delta(\mathbf{x},z)$.  If the density field grows non-linearly
but the velocity field still remains in the linear regime, then the initial Gaussian 
density field is described by a log-normal density distribution 
\citep{coles.jones1991}. 
This log-normal model has been shown to describe the matter distribution in
hydrodynamical simulations \citep{bi.davidsen1997} and to reproduce the
observations of the IGM \citep{choudhury2011}.
For a log-normal density field, the baryon number density is given by
\begin{equation}
n_B(\mathbf{x},z)=n_0(z)e^{-\Delta_B^2(z)/2} e^{\delta_B(\mathbf{x},z)}
\label{eq:n_B}
\end{equation}
where $\delta_B$ is the linear baryon density field and 
$\Delta_B^2=\langle\delta_B^2\rangle$ its variance, with
\begin{equation}\label{sigma}
\Delta_B^2(z)=\langle \delta_B^2(\mathbf{x},z)\rangle 
= D^2_+(z)\int \frac{P_{DM}(k)}{[1+k^2L_0^2]^2}\frac{d^3k}{(2\pi)^3}
\end{equation}
If $\delta_B$ is Gaussian distributed, then the spatial mean 
$\langle n_B(\mathbf{x},z)\rangle$ equals the cosmic baryonic background 
density $n_0(z)$. Hereafter we shall denote by $\xi=n_B(\mathbf{x},z)/n_0(z)$
the non-linear baryon density in units of the baryon mean density.

In \cite{atrio.muecket2006}, \cite{atrio.muecket.santos2008} and 
\cite{genova.atrio.muecket2009} we assumed that the gas in the WHIM follows a 
polytropic equation of state. High resolution hydro-simulations have shown
that at the physical conditions of the WHIM, the shock heated gas
at temperatures $10^5 K < T_e \le 10^7 K$ and density contrasts $ 1\le\xi\le 100$
is not polytropic. \cite{kang.ryu.cen.song2005} and \cite{cen.ostriker2006}
provide phase diagrams of electron temperature $T_e$ versus baryon 
density $n_B$ in the range of interest.
Since the IGM is highly ionized due photo-ionization by the UV background, then
$n_e\approx n_B$. This is even more accurate for the WHIM as collisional ionization 
at high temperatures also contributes to maintain $n_e = n_B$. 
We have used these results to construct fits to the baryon temperature-
density distribution, $T_e = T_e(n_e)$. 
We used the fitting formula
\begin{equation}
\label{temp}
\log_{10}\left(\frac{T_e(\xi)}{10^8{\rm K}}\right) = 
- \frac{2}{\log_{10}({4 + \xi^{\alpha+ 1/\xi}})}\,.
\label{eq:temperature}
\end{equation}
This expression fits the results of \cite{kang.ryu.cen.song2005} when $\alpha=1-4$. 

The phase space derived from simulations are not simple linear relations
of the type $T_e=T_e(n_e)$; phase diagrams have a large scatter.
This theoretical uncertainty is important since the Comptonization parameter
and temperature anisotropies depend on how steeply the temperature
rises with increasing density contrast. We model this uncertainty
by taking $\alpha=1-4$ in eq.~(\ref{eq:temperature}), a range wide enough 
to reproduce sufficiently well the phase diagrams and the variation therein.
We have also considered the results of \cite{cen.ostriker2006}; 
after correcting a possible typo their 
expression reads $\log_{10}(T_e/10^8{\rm K})= -2.5/\log_{10}{(4+\xi)^{0.9}}$.
The phase diagrams obtained from cosmological simulations agree qualitatively
with the results of \cite{klar.muecket2012} on the effect of shocks on WHIM 
filaments. In particular, as time evolves shocks propagate to lower 
density regions, steepening the $T_e-n_e$ relation as shown in the phase diagrams 
of numerical simulations.

\begin{figure}
\includegraphics[width=60mm,angle=-90]{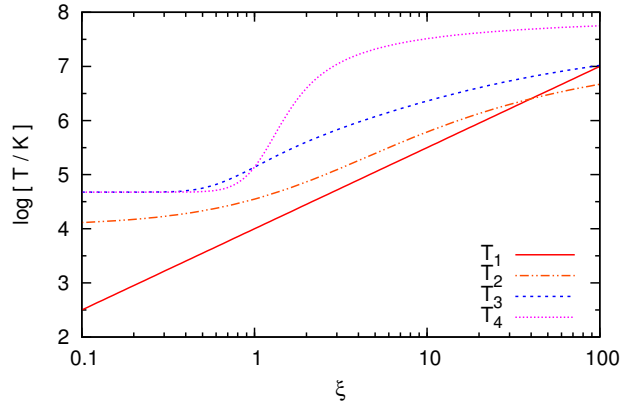}
\caption{Various fit functions $T_e=T_e(n_e)$ to phase diagrams 
derived from hydro-numerical simulations. Curves $T_3,\,T_4$ correspond
to the results of Kang et al. (2004) with $\alpha=1.5,\,4$,
respectively; $T_2$ corresponds to Cen \& Ostriker (2006); finally,
$T_1$ represents the power law $T\propto n^{3/2}$.}
\label{T_fit}
\end{figure}	 
 
For illustration, in Fig.~\ref{T_fit} we plot a series of different fits.
The lines with labels $T_3$ and $T_4$ correspond to \cite{kang.ryu.cen.song2005}
model with $\alpha=1.5$ and $4$, respectively, $T_2$ corresponds to 
\cite{cen.ostriker2006} with the corrected formula given above. For comparison, 
$T_1$ corresponds to a simple power law with $T_e\propto n_e^{3/2}$.
From Fig.~\ref{T_fit}, we can immediately conclude that since
filaments of any density in the $T_3,\,T_4$ models
are hotter than in the $T_1,T_2$ models, temperature anisotropies
and distortions would be larger in the former than in the latter cases.

\section{The Comptonization parameter}\label{compton}

The log-normal formalism introduced in the previous section permits to compute 
the statistical properties of the TSZ generated by the baryon
population of WHIM filaments.  The average Comptonization parameter $y_{\mathrm{av}}$ 
along a {\it los} is \citep{atrio.muecket2006}
\begin{equation}\label{yparameter}
y_{\mathrm{av}}=y_0\int \langle n_e(z,x)T_e(z,x)\rangle dl\,, 
\label{eq:yaverage}
\end{equation}
where the electron and temperature distribution are given by eqs.~(\ref{eq:n_B}) and
(\ref{eq:temperature}), respectively.

The average in eq.~(\ref{eq:yaverage}) must be limited to those scales 
$(\xi_1,\xi_2)$ where the non-linear evolution is well described by the 
log-normal probability density function. The choice of $\xi_1$ is somewhat
arbitrary. Since only gas with density contrasts $\xi>2$ will undergo 
shock-heating, this could be taken as a lower limit. 
As a criteria, we take $\xi_1=\xi_{\mathrm{median}}+\sigma$, where $\sigma =\sqrt{var(\xi)} $ ($var(\xi)=\exp(\Delta_B^2) - 1$) is  the
standard deviation of the log-normal distribution .
To include only the scales that undergo shocks 
we also require that $\xi_1>2$ at all redshifts.
As upper limit we took $\xi_2\le 100$, but the results were weakly
dependent on this value.

Even though this criteria is somewhat arbitrary, the mass fraction on the 
interval $(\xi_1,\xi_2)$ is very close to that of the undetected baryons: 40-50\%.
At any redshift, the mass fraction of the gas contained in
filaments with overdensities in the range $[\xi_1,\xi_2]$ depends
on the IGM temperature. If $\bar{\xi}$ is the mean of the distribution then 
$M(\xi_1,\xi_2)=\bar{\xi}^{-1}\int_{\xi_1}^{\xi_2}\xi F(\xi)$
is the fraction of mass in the integration range. At $z=0$, the fraction of 
baryons in the WHIM is $\approx 48\%$ for $A=0.9$ and $\approx 43\%$ for $A=0.5$.
One would expect the mass fraction to be largest for $T_{IGM}=10^{3.6}$K,
since smaller scales would survive shock heating than for $T_{IGM}=10^4$K. 
The difference comes from the lower limit $\xi_1$. For $A=0.5$,
the rms of the log-normal distribution $\sigma$ is larger and 
so is $\xi_1$; the integration is carried over a smaller range 
of densities and low density regions that are included for $A=0.9$
are excluded for $A=0.5$. However, even if in this aspect the model is not 
fully satisfactory, the low dense regions do not contribute significantly to 
spectral distortions or temperature anisotropies and the exact value of the 
lower limit $\xi_1$, while important for the 
baryon fraction on the WHIM, has little effect on the final results.

\section{The angular power spectrum}\label{spower}

The correlation function of the CMB temperature anisotropies
due to the WHIM TSZ contribution along two different {\it los} $\mathbf{l}$ and 
$\mathbf{l'}$ separated by an angle $\theta$, 
is given by \citep{atrio.muecket2006}
\begin{equation}\label{corr}
C(\theta)=y_0^2 \int _0^{z_f}\int_0^{z'_f} 
\langle S(\mathbf{x},z)S(\mathbf{x'},z') \rangle \frac{dl}{dz} \frac{dl'}{dz'}dzdz'\,,
\end{equation}
where $S(\mathbf{x},z)=n_e(\mathbf{x},z)T(n_e(\mathbf{x},z))$. The density average 
of eq.~(\ref{corr}) is carried out over the same density interval than 
in eq.~(\ref{eq:yaverage}).

Since the equation of state of the gas is not longer polytropic,
the correlation of the electron density at different positions can
not be factored out. The average in eq.~(\ref{corr}) differs from our 
previous treatment since now it needs to be computed using the  
bi-variate log-normal PDF. Remembering that $\xi=n_B(\mathbf{x},z)/n_0(z)$, then
\begin{equation}
\langle S(\mathbf{x},z)S(\mathbf{x'},z') \rangle = 
\int \int \xi T(\xi) \xi'T(\xi')F(\xi,\xi') d\xi d\xi'\,,
\label{eq:ss}
\end{equation}
where the bi-variate log-normal PDF $F(\xi,\xi')$ is given by
 \begin{equation}
 \begin{array}{l}
  F(\xi,\xi')=\frac{1}{2\pi \,\xi\, \xi'\,\Delta_B\, \Delta_B' 
\sqrt{1-r_c^2}} \,\exp\Big[ -\frac{1}{2(1-r_c^2)}\times\nonumber\\ 
 \Big( \frac{(\log \xi - \mu)^2}{\Delta_B^2}
 -2r_c \frac{(\log \xi-\mu)(\log \xi'-\mu')}{\Delta_B\Delta_B'}+
\frac{(\log \xi'-\mu')^2}{\Delta_B'^2}\Bigg)\Bigg]\,.
 \end{array}
 \end{equation}
In this expression, $\Delta_B$ is given by eq.~(\ref{sigma}),
$\mu = -\Delta_B^2/2$ is the mean of the log-normal 
PDF and $r_c$ is the correlation coefficient
$r_c = \langle \log(\xi)\log(\xi')\rangle/(\Delta_B\Delta_B')$
that in this particular case is given by
\begin{equation}
r_c=\frac{D_+(z)D_+(z')}{2 \pi^2 \Delta_B \Delta_B'}
\int\frac{P_{DM}(k)\, 
j_0(k|\mathbf{x}-\mathbf{x}'|) k^2 dk}{[1+L_0(z')^2k^2][1+L_0(z)^2k^2]}\,.
\label{eq:q}
\end{equation}
In this expression, $j_0$ denotes the spherical Bessel function
of zero order and $|\mathbf{x}-\mathbf{x}'|$ the proper distance between two 
patches at positions $\mathbf{x}(z)$ and $\mathbf{x}(z')$ separated an angular
distance $\theta$. In the flat-sky approximation
\begin{equation}
|\mathbf{x}-\mathbf{x}'|\approx \sqrt{l_{\perp}(\theta,z)^2 + [r(z) -r(z')]^2} ,
\label{eq:x}
\end{equation}
with $l_\perp(\theta,z)$ the transverse distance of two points located 
at the same redshift.  Therefore, the angular dependence of the correlation
function in eq.~(\ref{corr}) enters only through eq.~(\ref{eq:x}) in eq.~(\ref{eq:q}).

The power spectrum can be obtained by the Fourier transform of the correlation 
function (eq.~\ref{corr})
\begin{equation}
C_\ell=2\pi \int_{-1}^{+1}C(\theta)P_\ell(cos(\theta))d(cos(\theta))
\label{eq:cl}
\end{equation}
where $P_\ell$ denotes the Legendre polynomial of multipoles $\ell$.

\section{Results}\label{results}

To compute the temperature anisotropies generated by the WHIM we take 
as a background cosmological model the concordance $\Lambda$CDM model
with densities dark energy, dark matter and baryon fractions
$\Omega_{\Lambda}=0.75$, $\Omega_m=0.25$, $\Omega_b = 0.043$ and 
Hubble constant $h=0.73$ and the amplitude of the matter density perturbations
at $8h^{-1}$Mpc, $\sigma_8=0.8$. For simplicity, we fix the frequency 
dependence of the TSZ effect to unity, $G(\nu)=1$. Our results depend 
also on physical parameters such as $L_0$ and the shape of
the equation of state. The former is determined by the mean
IGM temperature, characterized by a parameter $A\approx 0.5-0.9$ 
(see sec.~\ref{logmodel}) and for the latter we shall consider three 
different relations: 
A polytropic equation of state $T_e\propto n_e^{3/2}$, the Cen \& Ostriker (2006)
model and Kang et al (2004) model
given by eq.~(\ref{eq:yaverage}) with $\alpha=1.5,\, 4$, denoted by
by $T_1,\, T_2,\, T_3,\, T_4$, respectively.

\subsection{The mean Comptonization parameter}

\begin{figure}
\includegraphics[width=60mm,angle=-90]{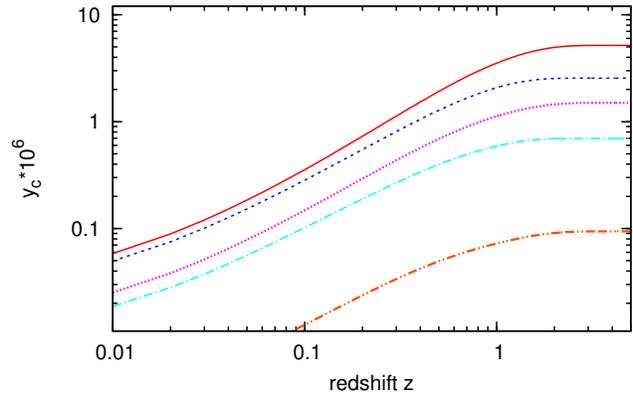}
\caption{Comptonization parameter $10^6y$ as a function of redshift.
From top to bottom  solid and dashed lines corresponds to model $T_4$
and $A=0.5,0.9$, dotted and dot-dashed lines to $T_3$ and $A=0.5,0.9$,
respectively. Finally the double dot-dashed line corresponds to $T_2$ with $A=0.5$.}
\label{y_par}
\end{figure}

In Fig. \ref{y_par} we plot the contribution to 
the average Comptonization parameter $y_{av}$ from $z=0$ up to the
given redshift for different models. In all cases,
most of the contribution comes from $z\le 1$. In retrospect,
this justifies restricting in sec.~\ref{logmodel} the IGM temperature parameter 
to the interval $A=0.5-0.9$, valid for the interval $z\le 3$, since
baryons at higher redshifts do not contribute to the TSZ effect.

In the figure, the amplitude of the CMB distortion depends on the equation
of state and the temperature of the IGM. When $T_{IGM}=10^{3.6}$K
smaller (and therefore more) scales contribute to the shock-heated WHIM and $L_0$ is smaller. Those small 
scales contribute to the variance of the baryon field at all redshifts 
(see eq.~\ref{sigma}). These regions are of high density and at
high temperature and the total effect increases, compensating for 
the baryon fraction being smaller (see section~\ref{compton}).

\begin{figure}
\includegraphics[width=60mm,angle=270]{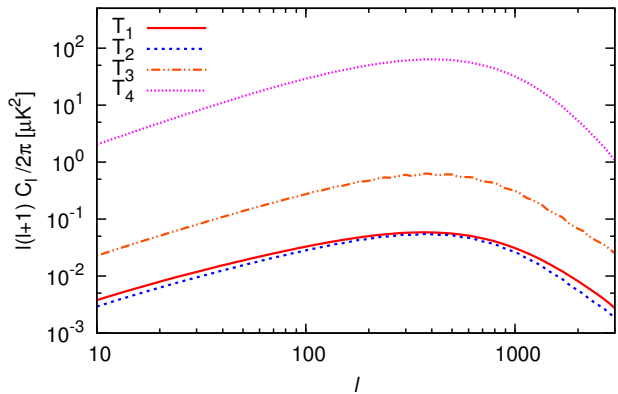}
\caption{Power spectrum of the radiation temperature anisotropies generated 
by the WHIM. Lines correspond to the models given in  Fig.\ref{T_fit}, with 
$A=0.5$.}
\label{power}
\end{figure}

\begin{figure}
\includegraphics[width=60mm,angle=270]{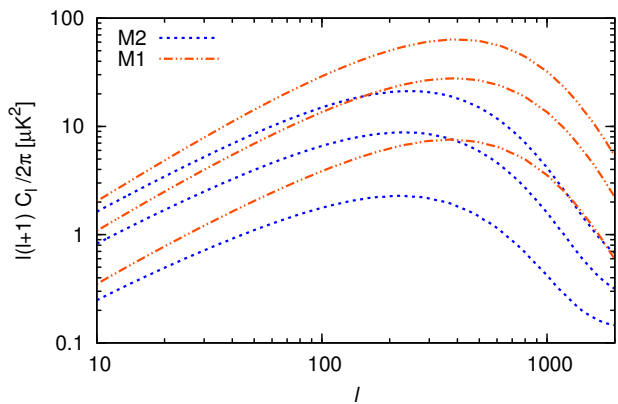}
\caption{Effect of the cut-off length in the radiation power spectra.
Curves $M_1,\, M_2$ correspond to $A=0.5,\, A=0.9$, respectively. From top
to bottom, the equation of state corresponds to the $T_4,\,T_3$ and $T_2$ models.}
\label{powert}
\end{figure}

\subsection{The correlation function and the power spectrum}\label{corrpower}

In Fig.~\ref{power} we show the power spectrum of the radiation temperature anisotropies
generated by the WHIM for different temperature models, obtained by numerical 
integration of eqs.~(\ref{corr}-\ref{eq:cl}). Lines follow the same convention 
than in Fig.~\ref{T_fit}. In Fig.~\ref{powert} we show the dependence with 
equation of state and $T_{IGM}$. The overall shape of the power spectra
are very simple: the functional form has a single maximum. 
For the range of variation of the model parameters, the maximum 
occurs in the range $\ell=200-500$ and the maximum amplitude is
$\ell(\ell+1)C_\ell/2\pi\simeq 0.07 - 70 (\mu K)^2$.  
The different equations of state $T_1-T_4$ give
power spectra that differ by three orders of magnitude in amplitude. 
This wide range reflects our
theoretical uncertainty on the equation of state of the shock-heated gas.
Using a polytropic equation of state gave even a larger range, and a
better determination of the TSZ power spectrum can only come
through better understanding of the physics of the WHIM, that is, through 
cosmological hydrosimulations with larger dynamical range.

The effect of the IGM temperature changes both the amplitude and the location of the
maximum. This was to be expected since for smaller IGM temperature
the cut-off length $L_0$ is smaller and the contribution of small scales is 
more important, increasing the amplitude reducing the average angular size 
of the anisotropies. For $A=0.5$ the power spectrum reaches a maximum at 
$\ell\simeq 450$ with an amplitude of $\ell(\ell+1)C_{\ell, \mathrm{max}}/2\pi 
\approx 70 (\mu K)^2$.  For $A=0.9$,
the maximum is at $\ell \sim 300$ and its amplitude is smaller
by a factor $\sim 3$.

\begin{figure}
\includegraphics[width=60mm,angle=-90]{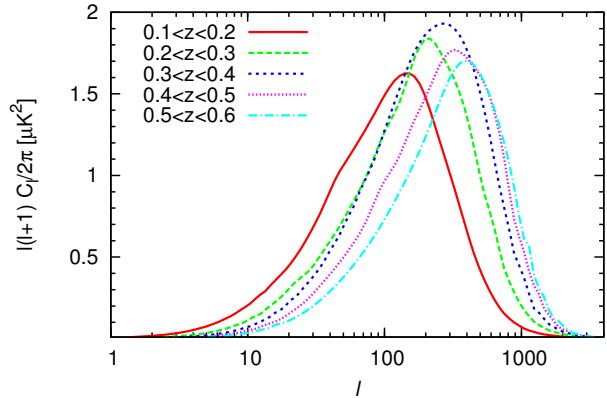}
\caption{Contribution to the power spectrum of different
redshift intervals, for the $T_4$ model with $A=0.5$.}
\label{zdependence}
\end{figure}

In Fig. \ref{zdependence} we show the contribution to the power 
spectrum due to the WHIM at different redshift intervals. Similarly to what
occurs with the Comptonization parameter, the dominant
contribution comes $0.3<z<0.5$, and the contribution for structures
with $z>1$ is negligible, in agreement with the results of numerical
simulations reported by \cite{smith.et.al2011}. Again, this justifies
the use of a simple parametrization of the IGM temperature that
is only valid for redshifts $z\le 3$.

\begin{figure}
\includegraphics[width=60mm,angle=270]{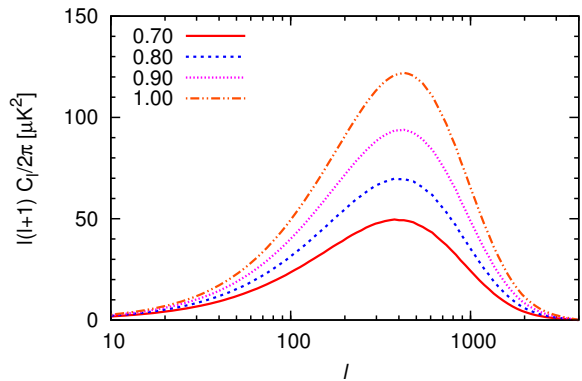}
\caption{Radiation power spectrum for the $T_4$ model with
$A=0.5$ and different values of $\sigma_8$.}
\label{power_s8}
\end{figure}

The radiation power spectrum of the WHIM induced temperature anisotropies 
also depends on the cosmological parameters of the concordance model.
As the amplitude of the TSZ effect depends on the number of electrons 
projected along the $los$, the amplitude of the effect scales 
as $\Omega_Bh$. Other parameters like $\Omega_m$ or $\Omega_\Lambda$
change the induced temperature anisotropies because they change the shape
matter power spectrum in eq~.(\ref{sigma}). But the variance of the
linear baryon density field $\Delta_B^2$ is dominated by the amplitude
of the matter power spectrum at small scales, and the largest effect is
produced by variations in the amplitude at those scales. Then,
the largest variations are produced by changes in $\sigma_8$. These are
shown in Fig.~\ref{power_s8}. A linear regression fit to the power spectra 
of different cosmological parameters gives the following scaling relation
at the maximum
\begin{equation}
\label{eq:scaling}
[(\ell+1)\ell C_\ell]_{\mathrm{max}}/2\pi \propto \sigma_8^{2.6}(\Omega_b h)^2.
\end{equation}
This scaling relations differ from our previous results in two very important
aspects: the power spectrum does not depend so strongly on the model parameters
and the TSZ contribution of the WHIM is always smaller than that of clusters.
However, like in Atrio-Barandela \& M\"ucket (2006), the bulk of the contribution
occurs well within the angular range measured by WMAP and Planck. This opens
the possibility of looking for the WHIM contribution in the radiation
power spectrum as in G\'enova-Santos et al (2009). A more suitable data
set is the forthcoming Planck data. The high resolution and wide frequency
coverage of this instrument makes it feasible to trace the WHIM
not only by its power spectrum, but also by the changing amplitude with
frequency. In particular, the 217GHz channel, that corresponds to the 
TSZ null, will help to separate the signal from foreground
contributions and other larger systematics. 

\section{Conclusions}\label{conclusion}

We have shown that the gas within non-linear large-scale filaments,
with density contrasts $2\le \xi \le 100$, shock-heated up to
temperatures $10^5 K< T < 10^7 K$ will produced measurable temperature
anisotropies via the TSZ effect. We have assumed that the gas follows
the DM distribution that is well described by a log-normal PDF 
in the density range of interest, in agreement with 
N-body simulations. The main difference with 
Atrio-Barandela (2006) and Atrio-Barandela et al (2008) is the
adoption of an equation of state $T_e=T_e(n_e)$,
appropriate to describe the shock-heated gas in the WHIM.
The equation of state was derived from phase-diagrams of cosmological
hydrodynamical simulations. This complicated the statistical treatment,
requiring the statistical averages over densities to be carried out using
the bi-variate log-normal PDF. 
Our final results depend on the cosmological models, the temperature
of the IGM and the functional form of the equation of state. We showed
that for temperatures compatible with observations and for functional
relations derived from numerical simulations, the amplitude
of the radiation power spectrum is $\ell(\ell+1)C_\ell/2\pi=0.07-70 (\mu K)^2$,
being its maximum amplitude at $\ell\simeq 200-500$ and with the bulk of the
contribution coming from redshift $z\le 1.0$. The large scatter in the
final values of the Comptonization parameter and in the
amplitude of the power spectra reflects the theoretical uncertainty
associated with the functional shape of the temperature-density relation
of the shock-heated gas.

The WHIM contribution peaks at a much larger angular scale than 
$\ell\sim 3000$, that of clusters 
of galaxies, making it accessible to the observations of WMAP and Planck.
Recently we have analyzed \citep{suarez2012} the observational
prospects of the WHIM TSZ by cross-correlating CMB data with tracers
of the gas distribution. We concluded that for the current model parameters,
the density field reconstructed from the 2MASS galaxy catalog and Planck
data could provide the first evidence of the large scale distribution of
the WHIM. 

\section*{Acknowledgments}

ISV thanks the DAAD for the financial support, grant A/08/73458.
FAB acknowledges financial support from the Spanish
Ministerio de Educaci\'on y Ciencia (grants FIS2009-07238
and CSD 2007-00050). He also thanks the hospitality of
the Leibniz-Institut f\"ur Astrophysik Potsdam.

\bibliographystyle{mn2e}
\bibliography{mn-jour,bibliografia}	
\label{lastpage}
\end{document}